\documentclass[twocolumn,showpacs,amsmath,amssymb]{revtex4}
\usepackage{latexsym}
\usepackage{pifont}
\usepackage{color}
\usepackage{epsfig}

\newcommand{\nn}{\nonumber\\&&}

\newcommand{\bfr}{{\bf r}}

\newcommand{\ben}{\begin{displaymath}}
\newcommand{\een}{\end{displaymath}}
\newcommand{\be}{\begin{equation}}
\newcommand{\ee}{\end{equation}}
\newcommand{\bea}{\begin{eqnarray}}
\newcommand{\eea}{\end{eqnarray}}

\newcommand{\eq}[1]{Eq.~(\ref{#1})}

\begin{document} 
\title{\bf % revision March 5}
Natural  Resolution of the Proton Size Puzzle} 
\author{G.~A.~Miller,$^1$ A.~W.~Thomas$^2$, J.~D.~Carroll$^2$, and J. Rafelski$^3$}
\affiliation{$^1$ Department of Physics, University of Washington,  Seattle, WA 98195-1560,\\
$^2$
CSSM, School of Physics and Chemistry, University of Adelaide, Adelaide SA 5005, Australia \\
$^3$ Department of Physics, University of Arizona, Tucson, Arizona 85721, USA }
\date{January 21, 2011}
\begin{abstract}
{We show that off-mass-shell effects arising from the internal structure of the proton provide 
a new proton polarization mechanism in the Lamb shift, proportional to  the lepton mass to the fourth power.
This effect is  
capable of resolving the current puzzle regarding the difference in the proton radius 
extracted from muonic compared with electronic hydrogen experiments.  These off-mass-shell effects
could be probed in several other experiments. 
}
\end{abstract}

\maketitle
 
The recent, extremely precise extraction of the proton radius~\cite{pohl} from the measured 
energy difference between the $2P_{3/2}^{F=2}$ and  $2S_{1/2}^{F=1}$ states of muonic hydrogen (H) 
has created considerable interest.  Their analysis yields a proton radius that is smaller than 
the CODATA~\cite{codata} value (extracted mainly from electronic H) by about 4\% or 5.0 
standard deviations.  This implies~\cite{pohl} that either the Rydberg constant has to be 
shifted by 4.9 standard deviations or that 
the QED calculations for hydrogen are insufficient. 
Since the Rydberg constant is extremely well measured, and the 
QED calculations seem to be very extensive and highly accurate, the muonic H finding presents 
a significant puzzle to the entire physics community.
 
Our analysis is motivated by the fact that muonic hydrogen is far smaller than electronic hydrogen 
and therefore more sensitive to corrections arising from hadron structure. In particular, we 
consider the lowest order correction associated with off-shell behaviour at the photon-nucleon 
vertex, showing that it can very naturally account for the difference reported by Pohl {\it et
al.}. Since at the present state of development of hadronic physics it is not possible to 
provide a precise value for this correction, our result may be viewed as a phenomenological 
study of the sensitivity of muonic hydrogen to important aspects of proton structure. It should 
spur further study of processes which could be sensitive to off-shell changes in proton 
structure. In alternate language, the explanation which we present may be viewed 
as a new contribution from proton polarization that is not constrained by 
dispersion relations but which can be studied in systems other than the hydrogen atom. 

We begin with a brief discussion of the relevant phenomenology. Pohl {\it et al.} show that 
the energy difference
between the  $2P_{3/2}^{F=2}$ and  $2S_{1/2}^{F=1}$ states, $\Delta\widetilde{E}$   is given by
\bea 
\Delta\widetilde{E}=209.9779(49)-5.2262r_p^2+0.0347 r_p^3  \;{\rm meV},\label{rad}
\eea
where $r_p$ is given in units of fm. Each of the three coefficients is obtained from extensive 
theoretical work ~\cite{eides,friar,atomcalcs,zemach,grotchyennie}, typically  confirmed by several groups. 
Studies of the relevant atomic structure calculations and 
corresponding efforts to 
improve those have revealed no variations large enough to significantly affect the above 
equation~\cite{jent,carroll}. Using  
this equation, we see that the difference between the Pohl and CODATA values of the proton radius  
would be entirely removed by an increase of the first term on the rhs of Eq.~(1) 
%%%first change
by just 0.31 meV=$3.1\times 10^{-10}$  MeV, but an effect of even half that much would
be large enough to dissipate the puzzle. 
Finding a new effect of about that value 
%%% second change
 resolves the puzzle provided that the 
corresponding effect  in electronic H is no more than a few  parts in a million 
(the  current difference between theory and experiment~\cite{eides}). An effect that gives a 
contribution to  $\Delta\widetilde{E}$  of the form $\alpha^5 {m^4\over M^3}$ 
(with $m$ the lepton mass and $M$ the proton mass) could therefore resolve 
the proton radius puzzle and cause no disagreement in electronic  H.

The search to find such an effect has attracted considerable interest. New physics beyond  
the Standard Model  must satisfy a variety of low-energy constraints and so far no explanation 
of the proton radius puzzle has been found that satisfies these  
constraints~\cite{Barger:2010aj,Jaeckel:2010xx,TuckerSmith:2010ra,jent1,Brax:2010gp}.  
Attention has been paid to the third term of \eq{rad}~\cite{dr}, with the result that its current 
uncertainties are far too small to resolve the proton radius puzzle~\cite{Cloet:2010qa,walcher}.

\newcommand {\boldsigma}{\mbox{\boldmath$\sigma$}}
\newcommand {\boldgamma}{\mbox{\boldmath$\gamma$}}
\newcommand {\boldSigma}{\mbox{\boldmath$\Sigma$}}

We therefore seek an explanation based on the fact that the proton 
is not an elementary Dirac particle, 
and that many features of its interactions are still unknown. In particular, consider 
the electromagnetic 
%%%several changes here
vertex function %form factor, 
which  must depend on all of the relevant invariants.  
For a proton of initial 
four-momentum $p$, the most general expression %for the electromagnetic  form factor 
must include a term, dependent on the proton virtuality, that is 
proportional to $p^2-M^2$ and/or  ${p}\cdot\gamma_N-M$, where the subscript $N$ denotes acting on a nucleon, and 
$M$ is the nucleon mass.  Such terms have been discussed for a very  long time 
in atomic~\cite{zemach,grotchyennie}
and nuclear physics~\cite{emc}-\cite{Cloet:2009tx}. They have been of special concern in relation 
to the  difference between free and bound deep inelastic structure functions measured in 
the EMC effect~\cite{emc}-\cite{Ciofi}, %which corresponds roughly to a 10\%-15\% effect,
nucleon-nucleon scattering~\cite{Gross:2008ps} 
 and electromagnetic interactions involving 
nucleons~\cite{bincer,Naus:1987kv}, notably quasi-elastic
scattering~\cite{bobm}-%Strauch:2010nm,Paolone:2010qc,Lu:1997mu,Lu:1998tn,
\cite{Cloet:2009tx}.

Many  possible forms 
\cite{bincer,Naus:1987kv}  include the effects of proton virtuality; we consider three that
could be significant for the Lamb shift.
We write 
 the Dirac part of the  vertex function
 for a proton of momentum $p$ to absorb a photon of momentum $q=p'-p$.as:
%%%many changes in this paragraph 
%
\bea 
&&\hspace*{-0.5cm}\Gamma^\mu(p',p)= 
   \gamma_N^\mu F_1(-q^2) + %{\lambda\over M}
    F_1(-q^2)F(-q^2){\cal O}_{a,b,c}^\mu\label{off} \\
&&\hspace*{-0.5cm}{\cal O}_a^\mu={(p+p')^\mu\over 2M}[\Lambda_+(p') {(p\cdot\gamma_N -M)\over M}+{({p}'\cdot\gamma_N-M)\over M} \Lambda_+(p)]\nn
\hspace*{-0.5cm}{\cal O}_b^\mu=%\left[
((p^2-M^2)/M^2+({p'}^2-M^2)/M^2)\gamma_N^\mu\nn
\nn
\hspace*{-0.5cm}{\cal O}_c^\mu=
\Lambda_+(p')\gamma_N^\mu {(p\cdot\gamma_N -M)\over M}+
{({p}'\cdot\gamma_N-M)\over M} \gamma_N^\mu\Lambda_+(p)\nonumber
%\right]
,%\nonumber\\
\eea
where  three possible forms are displayed. Other terms of the vertex function needed to satisfy the WT identity do not contribute significantly  to the Lamb shift and are not shown explicitly.
 The proton Dirac form factor, $F_1(-q^2)$ is   empirically well represented as a dipole
$F_1(-q^2)=(1-q^2/\Lambda^2)^{-2},$  ($\Lambda $ = 840 MeV) for %. This gives a good representation of the data for 
the values of $-q^2\equiv  Q^2>0$ of up to about 1 GeV$^2$  needed here. %The second term on the right hand side of \eq{off}
%is a natural off-shell extension where
%the dimensionless parameter $\lambda$ represents the strength of the effect, 
$F(-q^2)$ is an off-shell form factor, and $\Lambda_+(p)=(p\cdot\gamma_N+M)/(2M) $ is an operator that projects on the on-mass-shell proton state.  We use  ${\cal O}_a$ unless otherwise stated. 
%The virtuality of the proton is represented  by  terms involving  $p\cdot\gamma_N-M$ or $p^2-M^2. $

We take the off-shell form factor $F(-q^2)$  to  vanish at $q^2=0$. This means that the charge of the off-shell proton will be the same as the charge of a free  proton, and is demanded by current conservation as expressed through the Ward-Takahashi identity \cite{bincer,Naus:1987kv}. %Otherwise, it is completely unknown. 
We assume
\bea 
F(-q^2)={{-\lambda q^2/ b^2}\over(1-q^2/\widetilde{\Lambda}^2)^{1+\xi}} .
\label{model2}
\eea
This purely phenomenological form is simple and   clearly not unique. 
The parameter $b$ is expected to be of the order of the pion mass, because these longest 
range components of the nucleon are least bound and more susceptible to the external perturbations 
putting the nucleon off its mass shell. At large values of $|q^2|$, $F$ has the  same  fall-off as $F_1$, 
if $\xi=0$.
%But there is no reason why $\xi$ should vanish.
 We take $\widetilde{\Lambda}=\Lambda$ here.

We briefly discuss the expected influence of using \eq{off}.
The ratio, $R$, of  off-shell effects  to on-shell effects, $R\sim\frac{(p\cdot\gamma_N-M)}{M} \lambda \frac{q^2}{b^2},\;(|q^2|\ll\Lambda^2)$
 is constrained by  a variety of nuclear phenomena such as the EMC effect (10-15\%), uncertainties in quasi-elastic electron-nuclear 
scattering~\cite{bobm}, and deviations from the Coulomb sum rule~\cite{zm}. For a nucleon experiencing a 50 MeV central potential,
$(p\cdot\gamma_N-M)/M\sim 0.05$, so $\lambda q^2/b^2$ is of order 2. 
%Typical effects at the center of nuclear matter are constrained 
%Such effects are typically constrained to be no larger than 10-15\%. 
%The relevant quantity determining  the influence of off-shell effects 
%is the ratio, $R$
%\bea R=0.04 \;\lambda\; F(-q^2).\label{ratio}\eea
%Thus existing nuclear physics knowledge prevents the magnitude of $\lambda$ from being 
%chosen arbitrarily large. 
The nucleon wave functions of 
light-front quark-models~ \cite{Lepage:1982gd}  contain  a propagator depending on $M^2$. Thus   
% of the form $D(M^2)\equiv M^2- \sum_i{\bfk_{\perp,i}+m_i^2\over x_i}$. T
the effect of nucleon virtuality is proportional to the derivative of the propagator with respect to $M$, or of the order of the wave function divided by difference between quark kinetic energy  and $M$. This is about three times the average momentum of a quark $(\sim 200 $ MeV/c) divided by the nucleon radius or roughly $M/2$. Thus $R\sim (p\cdot\gamma_N-M)2/M$,
% ${\partial \psi\over \partial M^2} \sim {1\over D(M^2) }\psi$. The average value of the 
%denominator is typically of the order $M^2$ 
and
the natural value of   $\lambda q^2/b^2$  is of order 2.

\begin{figure}[tb]
\epsfig{file=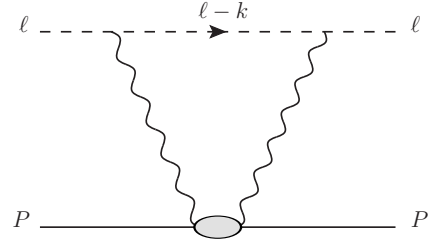%s2g.pdf
, width=6.0cm}\caption{Direct two-photon exchange graph corresponding to the
  hitherto neglected term. The dashed
  line denotes the lepton; the solid line, the nucleon; the wavy lines photons; and the ellipse the off-shell nucleon.}
 \label{fd1} \end{figure}

The lowest order term in which the nucleon is sufficiently off-shell in a muonic atom for this 
correction to produce a significant effect is the two-photon exchange diagram of 
 Fig.~\ref{fd1} and its crossed partner, including  an interference between one on-shell and one off-shell part of the vertex function.
  %evaluated in the lepton/nucleon rest frame  
%%% changed equation here
 The change in the invariant amplitude, $ {\cal M}_{\rm Off}$, due to using \eq{off}  along with ${\cal O}^\mu_a$, to be evaluated between fermion spinors,
 is given in the rest frame by
 %to be evaluated in the rest-frame between fermion spinors is 
\bea
&&\hspace*{-0.4cm} {\cal M}_{\rm Off}=
{ e^4 \over 2M^2}\int{ d^4k\over (2\pi)^4}{F_1^2(-k^2)F(-k^2)\over(k^2+i\epsilon)^2}  \label{md}\\
&&\times(  \gamma_N^\mu(2p+k)^\nu +\gamma_N^\nu(2p+k)^\mu)\nonumber\\
%\times\nonumber\\
&&\hspace*{-0.4cm}\times\left[\gamma_\mu{(l\cdot\gamma-k\cdot\gamma+m)
   \over k^2-2l\cdot k+i\epsilon}\gamma_\nu+
\gamma_\nu{(l\cdot\gamma+k\cdot\gamma+m)\over k^2+2l\cdot k+i\epsilon}\gamma_\mu\right],\nonumber%\nonumber
%({1\over k^2+i\epsilon})^2\nonumber\\&&\times{\gamma_N^\mu}%{( P\cdot\gamma+k\cdot\gamma+M)}
\eea
where  the lepton momentum is $l=(m,0,0,0)$, %and $m$ its mass, 
the virtual photon momentum is $k$ 
and the nucleon momentum $p=(M,0,0,0)$.  %Dirac operators with no subscript act  on the electron.
The intermediate  proton propagator is cancelled by the off-mass-shell terms of \eq{off}. 
This graph can be thought of as involving a contact interaction and the amplitude 
in \eq{md}  
as a new  proton polarization correction corresponding to a subtraction term in the 
dispersion relation for the two-photon exchange diagram that is not constrained by the  
cross section data~\cite{drell}. The resulting
virtual-photon-proton Compton scattering amplitude, containing the operator $\gamma^\mu_N\gamma^\nu_N$  corresponds to the $T_2$ term of conventional notation~\cite{Bernabeu:1973zn},
\cite{kp99}.
 \eq{md} is gauge-invariant;  not changed by adding  a term of
the form $k^\mu\;k^\nu/k^4$ to the photon propagator.

Evaluation proceeds in a standard way by taking the sum over Dirac indices, performing the 
integral over $k^0$ by contour rotation, $k^0\rightarrow-ik^0$, and integrating over the angular variables.  
The matrix element ${\cal M}$ is well approximated by a constant in momentum space, for  
momenta typical of a muonic atom, 
and  the corresponding potential
$V=i{\cal M}$  has the form $V(\bfr)=V_0\delta(\bfr)$  in coordinate space. 
This is  the ``scattering approximation"~\cite{eides}. 
Then the relevant matrix elements have the form $V_0\left|\Psi_{2S}(0)\right|^2$, 
where $\Psi_{2S}$ is the muonic hydrogen wave function of  the state  relevant to the 
experiment of Pohl {\it et al.} %This potential gives no contribution to $P$ states.
We use $
\left|\Psi_{2S}(0)\right|^2=(\alpha m_r)^3/ (8\pi)$, with 
the lepton-proton reduced mass, $m_r$.
The result 
\bea 
\langle 2S|V|2S\rangle&=&{-\alpha^5m_r^3\over M^2}{8\over\pi}\lambda {mM\over b^2}%\left[
F_L(m)%+\boldsigma\cdot\boldsigma_N F_{HFS}(m)\right]
\;,\notag\\[0.2cm]
F_L(m)&\equiv&{1\over \beta}\int_0^\infty 
{{ \sqrt{{(x+\beta )}/{x}}-1 }\over (1+x^2)^{5+\xi}} { \,x dx}\;,\label{fl}%\\[0.2cm]
%\tan^{-1}(x^2{\Lambda^2\over b^2})\;
%\frac
%{\beta },
%F_{HFS}(m)&\equiv&-{2\over\beta}\int_0^\infty {{1-\sqrt{{ (x+\beta )/\beta}}}\over (1+x^2)^{5+\xi}}\, x dx\;,\label{fhfs2}
\eea 
(where $\beta\equiv 4m^2/\Lambda^2$) shows
a new contribution to the Lamb shift, proportional to  $m^4$ and therefore negligible for electronic hydrogen.  Using  ${\cal O}_a^\mu$ leads to a vanishing hyperfine HFS splitting because the operator $\gamma^\mu_N$ is odd unless $\mu=0$.

We next seek  values of the model parameters $\lambda,b ,\xi$  of \eq{model2},
 chosen 
to reproduce the value of the needed energy shift of 0.31 meV with a value of $\lambda$ of order unity. 
Numerical evaluation,  using $\xi=0,\;\widetilde{\Lambda}=\Lambda$, 
 shows that 
 \bea \frac{\lambda}{b^2}=\frac{2}{(79\mathrm{MeV})^2}\label{res} \eea
  leads to 0.31 meV.  If $\xi$ is changed substantially from 0 to 1 %(using $\tilde{\Lambda}=\Lambda$)
the required value of $\lambda$% required to reproduce a 0.31 meV change in the Lamb shift 
would be  increased by about 10\%.
If our mechanism increases the muonic Lamb shift by 0.31 meV, the  change in the electronic H Lamb shift for the 2S-state is about 9 Hz, significantly below the current uncertainty in both theory and experiment~\cite{eides}.  
%The corresponding  HFS shift in  electronic H  is about 20 Hz,  again within current experimental and theoretical limits. 
Should some other effect account for part of the proton radius puzzle, the value of $\lambda/b^2$ would be  decreased. We also caution that other systems in which one might aim to test this 
effect could show sensitivity to the value of $\xi$ or $\tilde{\Lambda}$ in Eq.~(\ref{model2}).
%Using  ${\cal O}_{b,c}^\mu$ would lead to similar numerical results for the Lamb shift.

The other operators appearing in \eq{off} yield similar results when used to evaluate ${\cal M}_{\rm off}$.
 Using ${\cal O}_{b}$, gives  a term of the $T_2$ form with  a Lamb shift twice
  that of  ${\cal O}_{a}$, and also a HFS term that is about -1/6 of its Lamb shift, so the value of $\lambda/b^2$ would 
  be decreased  by 3/5. The use of  ${\cal O}_{c}$, gives a term of the $T_1$ form and  the same Lamb shift as  ${\cal O}_{a}$, as well as  a HFS term that is -1.7 times its Lamb shift. In this case, the value of $\lambda/b^2$ would be about -~3/2 times that of \eq{res}. The HFSs are  small enough to be well within current experimental and theoretical limits for electronic hydrogen. Thus  each operator leads to a reasonable explanation of the proton radius puzzle.

%%%new para here
It is necessary to comment on the difference between our approach, which yields a relevant proton polarization effect, and  others \cite{kp99}, \cite{am05}  which do not. 
The latter  use a current-conserving representation of the virtual-photon proton scattering amplitude in
terms of two unmeasurable  scalar functions, $T_{1,2}$. Dispersion relations are used to relate $T_{1,2}$ to their measured imaginary parts. However, terms with  intermediate nucleon states are treated by evaluating  Feynman diagrams. This allows the removal of an infrared divergence by subtracting the first iteration of the effective  potential that appears in the wave function. But the Feynman diagrams involve off-shell nucleons, so that their evaluation for composite particles must be ambiguous.  For example, using  two different forms of the on-shell electromagnetic vertex function, related by using the Gordon identity, leads to results that differ.
  This ambiguity in obtaining $T_{1,2}$ 
is removed in our approach by postulating \eq{off} and evaluating its consequences.
Note
also that in order to evaluate the term involving $T_1$ using a dispersion
relation one must introduce a subtraction
 function, $T_1(0,q^2)$. This is  unconstrained by prior data 
 \cite{Bernabeu:1973zn} because the value of $\sigma_L/\sigma_T$ at $\nu=\infty$ is not determined~\cite{Abe:1998ym}.
   Pachucki~\cite{kp99},  in  Eq.(31),  assumes  a form proportional to $q^2$ (see our \eq{model2}) times the  very small proton magnetic polarizability. However we are aware of no published derivation of this result.

In conclusion, we have shown that a simple off-shell correction to the photon-proton vertex, which
arises naturally in quantum field theory and  
is of natural size and consistent with  gauge invariance, is capable of 
resolving the discrepancy between the extraction of the proton charge radius from 
Lamb shift measurements in muonic and electronic hydrogen. Off-shell effects of the proton form factor were  an explicit concern of 
both Zemach~\cite{zemach} and Grotch \& Yennie~\cite{grotchyennie}. 
However, it is only with the remarkable improvement in experimental precision recently achieved~\cite{pohl} that 
it has become of practical importance. Within the field of nuclear physics there is great interest in the role that the modification of nucleon 
structure in-medium may play in nuclear structure~\cite{Guichon:1995ue,Guichon:2004xg}.
 We stress that
the effect postulated here can be investigated in lepton-nucleus scattering via the binding effects 
of the nucleon, as well as by lepton-proton scattering in arenas where two photon (or $\gamma, Z$) 
effects are relevant.

\section*{\it Acknowledgments:}
This research was supported by the United States Department of Energy, (GAM) grant FG02-97ER41014; (JR) grant DE-FG02-04ER41318; (JDC, in part) contract DE-AC05-06OR23177 (under which Jefferson Science Associates, LLC, operates Jefferson
Lab), and by the Australian Research Council and the University of Adelaide (AWT, JDC). GAM and JR gratefully acknowledge the support and hospitality of the University of Adelaide while the project was undertaken. We thank M.~C.~Birse, J.~A.~McGovern, R.~Pohl,  and T. Walcher for useful discussions.

\end{document}